\documentclass[smallextended]{svjour3}       %
 \usepackage[final
 ]{changes}
\definechangesauthor[name={TMOTTM},color=red]{TT}
\usepackage{tikz}
\usepackage{tikz-timing}[2009/05/15]
\usepackage{graphicx}
\usepackage[utf8]{inputenc}
\usepackage{adjustbox}
\usepackage[toc,acronym]{glossaries}
\usepackage{subcaption}

\makeglossaries

\newacronym{AI}{AI}{Artificial Intelligence}
\newacronym{ANN}{ANN}{Artificial Neural Network}
\newacronym{ALU}{ALU}{Arithmetical and Logical Unit}
\newacronym{CPU}{CPU}{Central Processing Unit}
\newacronym{FPGA}{FPGA}{Field Programmable Gate Array}
\newacronym{FIFO}{FIFO}{First In/First Out storage}
\newacronym{HW}{HW}{hardware}
\newacronym{ISA}{ISA}{Instruction Set Architecture}
\newacronym{I/O}{I/O}{Input/Output}
\newacronym{HPC}{HPC}{Hight Performance Computing}
\newacronym{ICCB}{ICCB}{Inter-Core Communication Block}
\newacronym{LAN}{LAN}{Local Area Network}
\newacronym{MC}{MC}{Multi-Core and/or Many-Core}
\newacronym{MLP}{MLP}{Memory Level Parallelism}
\newacronym{OoO}{OoO}{Out-of-Order}
\newacronym{OS}{OS}{operating system}
\newacronym{PD}{PD}{Propagation Delay}
\newacronym{QT}{QT}{Quasi-Thread}
\newacronym{PU}{PU}{Processing Unit}
\newacronym{SPA}{SPA}{Single Processor Approach}
\newacronym{SW}{SW}{software}
\newacronym{HPL}{HPL}{High Performance Linpack}
\newacronym{HPCG}{HPCG}{High Performance Conjugate Gradients}

\newacronym{EMPA}{EMPA}{Explicitly Many-Processor Approach}
\newacronym{EPE}{EPE}{EMPA Processing Element}
\newacronym{EME}{EME}{EMPA Morphing Element}
\newacronym{ECE}{ECE}{EMPA Communicating Element}
\newacronym{EICB}{EICB}{EMPA Inter-Core Block}
\newacronym{ESME}{ESME}{EMPA Storage Manager Element}
\usepackage {xcolor}

\definecolor{webgreen}{rgb}{0,.5,0}
\definecolor{webbrown}{rgb}{.6,0,0}
\definecolor{webyellow}{rgb}{0.98,0.92,0.73}
\definecolor{webgray}{rgb}{.753,.753,.753}
\definecolor{webblue}{rgb}{0,0,.8}
\definecolor{webgreen}{rgb}{0, 0.5, 0} % less intense green
\definecolor{webred}{rgb}{0.5, 0, 0}   % less intense red
\begin{document}

\title{Which scaling rule applies\\
	to large Artificial Neural Networks\thanks{The paper is the extended version of the presentation held at the 22nd Int'l Conf on Artificial Intelligence (ICAI'20), Las Vegas, USA, as ICA2246.
	Project no. 136496 has been implemented with the support provided from the National Research, Development and Innovation Fund of Hungary, financed under the K funding scheme.
}

}
\subtitle{Technological limitations for biology-imitating computing}


		\author{J\'anos  V\'egh	}


\institute{Kalim\'anos BT, 
Debrecen, Hungary \\
              \email{janos.vegh@unideb.hu~ORCID:~0000-0002-3247-7810}           %  \\
}

\date{Received: date / Accepted: date}

\maketitle

\begin{abstract}
 
The experience shows that cooperating and communicating computing systems, comprising segregated single processors, have severe performance limitations, which cannot be explained using von Neumann's classic computing paradigm.  
In his classic "First Draft" he warned that using a "too fast processor" vitiates his simple "procedure" (but not his computing model!); furthermore, that using the classic computing paradigm for imitating neuronal operations, is \textit{unsound}.
Amdahl added that large machines, comprising many processors, have an inherent
	disadvantage. 
Given that \gls{ANN}'s components are heavily  communicating
with each other, they are built from \added{a large number of} components designed/fabricated for use in conventional computing, furthermore they attempt to mimic biological operation using improper technological solutions, their achievable payload computing performance
is conceptually modest. 
The type of workload that \gls{AI}-based systems generate leads to an exceptionally low payload computational performance, and their design/technology limits their size to just above the "toy" level systems:
the scaling of processor-based \gls{ANN} systems is strongly nonlinear.
\added[comment=R2/Q1]{Given the proliferation and growing size
of \gls{ANN} systems, we suggest ideas to estimate in advance 
the efficiency of the device or application.
The wealth of \gls{ANN} implementations, and the proprietary technical data do not enable more.}
Through analyzing published measurements we
provide evidence that the role of data transfer time drastically influences both \gls{ANN}s performance and feasibility.
It is discussed how some major theoretical limiting factors,  \gls{ANN}'s layer structure and their methods of technical implementation of communication affect their efficiency.
The paper starts from von Neumann's original model, without neglecting
the transfer time apart from processing time; derives an appropriate interpretation and handling
for Amdahl's law. It shows that, in that interpretation, Amdahl's Law correctly describes \gls{ANN}s.
\end{abstract}

	\section{Introduction\label{sec:Introduction}}
%	Single-processor performance stalled nearly two decades ago~\cite{GameOverYelick:2011}, mainly because of reaching the limits,
%	the laws of nature enable~\cite{LimitsOfLimits2014}. 
%	As we pointed out~\textbf{\cite{VeghTemporal:2020}}, one of the major reasons for stalling was tremendously extending the inherent idle waiting times in computing.

Given the proliferation of \gls{ANN}-based devices, applications, and methods, furthermore that even supercomputers are re-targeted for \gls{AI} applications,
the efficacy of such systems is gaining growing importance.
Von Neumann in his "First Draft"~\cite{EDVACreport1945}
provided an \textit{approximation},
for (the timing relations of) \textit{vacuum tubes only}. 
He warned (in his section 6.3), that because the data transfer time is neglected in his model, using a "too fast processor" vitiates the procedure; furthermore, that using his paradigm for imitating neuronal operations, is \textit{unsound}, given that the conduction (transfer) time is longer than the synaptic (processing) time.  This limitation means that in today's technology background, it is at least doubly unsound to apply his paradigm to describe scaling \gls{ANN}s\footnote{Given that the classic paradigm is unsound for describing neurons, their communicating network has not been tauched.}. 
However, von Neumann did not provide another procedure that can consider the case corresponding to the today's technology, and the case of neural computing, respectively~\textbf{\cite{VeghMissingSecondDraft:2020}}.
The question opens, how then \textit{networks of computing objects using neuronal operations}, mimicking or inspired by biology, \textit{can be scaled}.

From computational point of view, an \gls{ANN}
	is an adaptive distributed processor-based architecture widely used to utilize its inputs and simulate the human-processing in terms of computation, response, and decision-making.
To analyze their performance, we start from
the "first principles" of computing, scrutinizing the terms and omissions.
Amdahl's famous idea for distributed many-processor systems introduced "strong scaling". He predicted~\cite{ScalingParallel:1993}: "\textit{scaling thus put larger machines [the brain inspired computers built from
	components designed for \gls{SPA} computers] at an inherent
	disadvantage}".
It was guessed early~\cite{ScalingParallel:1993}, that the \textit{payload} performance of 
processor assemblies does not scale linearly with the number of processors. 
However, the appearance of  "massively parallel" systems improved the degree of parallelization so much that 
a new approximation: the "weak scaling"~\cite{Gustafson:1988} appeared.
Due to the peculiarities of its workload,
the \gls{AI}-related progress (including \gls{ANN}s)
shows up much worse scaling~\cite{DeepNeuralNetworkTraining:2016} than expected, and led to that "\textit{Core progress in AI has stalled in some fields}"~\cite{AIcoreProgressStalled:2020}.
Also, the Gordon Bell Prize jury noticed~\cite{GordonBellPrize:2017}
that "\textit{Surprisingly, there have been no brain
	inspired massively parallel specialized computers [among the winners]}".

All the reasons listed above have their root essentially in \textit{neglecting the temporal behavior of computing
	components and methods}.
In section \ref{sec:TheScaling} we shortly review the considered scaling methods, and some of their consequences.
In section \ref{sec:AmdahlsLaw}, Amdahl's idea is shortly described (and partly: reinterpreted): his famous formula using our notations is introduced.
In section \ref{sec:GustafsonLaw}, we scrutinize the primary purpose of massively parallel processing, Gustafson's idea.
Section~\ref{fig:AIperformanceLimit} discusses different factors
affecting computing performance of processor-based (as well as 
some aspects of other electronic equipments) \gls{ANN} systems.

	\section{Common scaling methods}
\label{sec:TheScaling}

The scaling methods used to model different 
implementations of parallelized sequential processing (aka "distributed computing") are 
approximations to the more general model presented in~\textbf{\cite{VeghTemporal:2020,VeghComputingModel:2021}}. 
As discussed in \textbf{\cite{VeghHowMany:2020,VeghModernParadigm:2019}}, \textit{parallelized sequential systems have their inherent performance limitation}.
Using that formalism and data from the TOP500 database~\cite{Top500:2016}, we could estimate performance limits for present supercomputers. It enabled us to comprehend why
supercomputers have their inherent performance limit~\textbf{\cite{VeghHowMany:2020}}.
For the accuracy of the scaling method, see the case of supercomputer Piz Daint in~\cite{VeghHowMany:2020}.
We also validated~\textbf{\cite{Vegh:2017:AlphaEff}}  our "time-aware scaling"
(as a mostly empirical experience) through applying it, among others,
for qualifying load balancing compiler, cloud operation, on-chip communication. Given that experts, with the same background, also build \gls{ANN} systems, from similar components, we can safely assume that the same scaling is valid for those systems, too.
Calibrating our systems for some specific workload
(due to the lack of validated data) is not always possible, but one can compare the behavior of systems and draw some general conclusions.

\subsection{Amdahl's Law\label{sec:AmdahlsLaw}}

Amdahl's Law (called also 'strong scaling') is usually formulated  as
\vspace{-.3\baselineskip}    
\begin{equation}
S^{-1}=(1-\alpha) +\alpha/N \label{eq:AmdahlBase}
\end{equation}

\noindent where $N$ is the number of parallelized code fragments, 
$\alpha$ is the ratio of parallelizable fraction to total (so $(1-\alpha)$ is the "serial percentage"),
$S$ is a measurable speedup.
That is, Amdahl's Law considers a \textit{fixed-size problem},
and $\alpha$ portion of the task is distributed to fellow processors.

When calculating the speedup, one calculates
\begin{equation}
S=\frac{(1-\alpha)+\alpha}{(1-\alpha)+\alpha/N} =\frac{N}{N\cdot(1-\alpha)+\alpha}
\end{equation}
However, as expressed in~\cite{AmdalVsGustafson96}: "\textit{Even though Amdahl's Law is theoretically correct, the serial percentage is not practically obtainable.}"
That is, concerning $S$, there is no doubt that it is derived as the ratio of
\textit{measured execution times}, for non-parallelized and parallelized cases, respectively.
But, what is the exact interpretation of $\alpha$, and how can it be used?

Amdahl listed performance affecting
factors, such as "boundaries are likely to
be irregular;
interiors are inhomogeneous;
computations required may be dependent on the states
of the variables at each point;
\textit{propagation rates of different physical effects may be quite different};
the rate of convergence or convergence at all may be strongly dependent on sweeping through the
array along different axes on succeeding passes, etc."
Amdahl has foreseen issues with "sparse" calculations
(or in general: \textit{the role of data transfer}) 
as well as that the \textit{physical size} of computer and the \textit{interconnection} of its computing units (especially in the case of distributed systems) also matters.

Amdahl used wording "\textit{the fraction of the computational load}", giving way to his followers to give meaning to that term. 
This (unfortunately formulated) phrase "\textit{has caused nearly three decades of confusion in the parallel processing community. This confusion disappears when processing times are used in the formulations}"~\cite{AmdalVsGustafson96}.
On one side, \textit{it was guessed that Amdahl's Law is valid only for software} (for the number of executed instructions), and on the other side \textit{other affecting factors, he mentioned but did not discuss in detail, were forgotten}.

Expressing Amdahl's speedup is not simple:  "\textit{For example, if the following percentage is to be derived from computational experiments, i.e., recording the total parallel elapsed time and the parallel-only elapsed time, then it can contain all overheads, such as communication, synchronization, input/output and memory access. The law offers no help to separate these factors. On the other hand, if we obtain the serial percentage by counting the number of total serial and parallel instructions in a program, then all other overheads are excluded. However, in this case, the
	predicted speedup may never agree with the experiments.}"~\cite{AmdalVsGustafson96}
Moreover, the experimental one is always smaller than the theoretical one.

From computational experiments, one can express $\alpha$ from Eq.~(\ref{eq:AmdahlBase}) in terms measurable experimentally as

\begin{equation}
\alpha = \frac{N}{N-1}\frac{S-1}{S} \label{equ:alphaeff}
\end{equation}

\noindent
It is useful to express \textit{ computing efficiency} with those experimentally measurable
\vspace{-.3\baselineskip}    
\begin{equation}
%\boxed{E(\large N,\alpha)} = 
E(\large N,\alpha) =
% \frac{S}{N}=\boxed{\frac{1}{\textcolor{red}{\Large N}\cdot(1-\alpha)+\alpha}}=
 \frac{S}{N}=\frac{1}{{\Large N}\cdot(1-\alpha)+\alpha}= \frac{R_{Max}}{R_{Peak}}
\label{eq:soverk}
\end{equation}
data. Efficiency is an especially valuable parameter, given that constructors of many parallelized sequential systems
(including TOP500 supercomputers) provide the efficiency (as $R_{Max}/R_{Peak}$) of their computing system, and of course, the number of processors $N$ in their system. Via reversing  Equ. (\ref{eq:soverk}), the value of $\alpha_{eff}$ can be expressed with measured data as
\begin{equation}
%\boxed{\alpha_{eff}(E,N)} = \boxed{\frac{E\cdot N -1}{E\cdot (N-1)}}\label{eq:alphafromr}
\alpha_{eff}(E,N) = \frac{E\cdot N -1}{E\cdot (N-1)}\label{eq:alphafromr}
\end{equation}

\begin{figure*}
	%    \hspace{-1cm}
	\includegraphics[width=\textwidth
	]{%fig/
		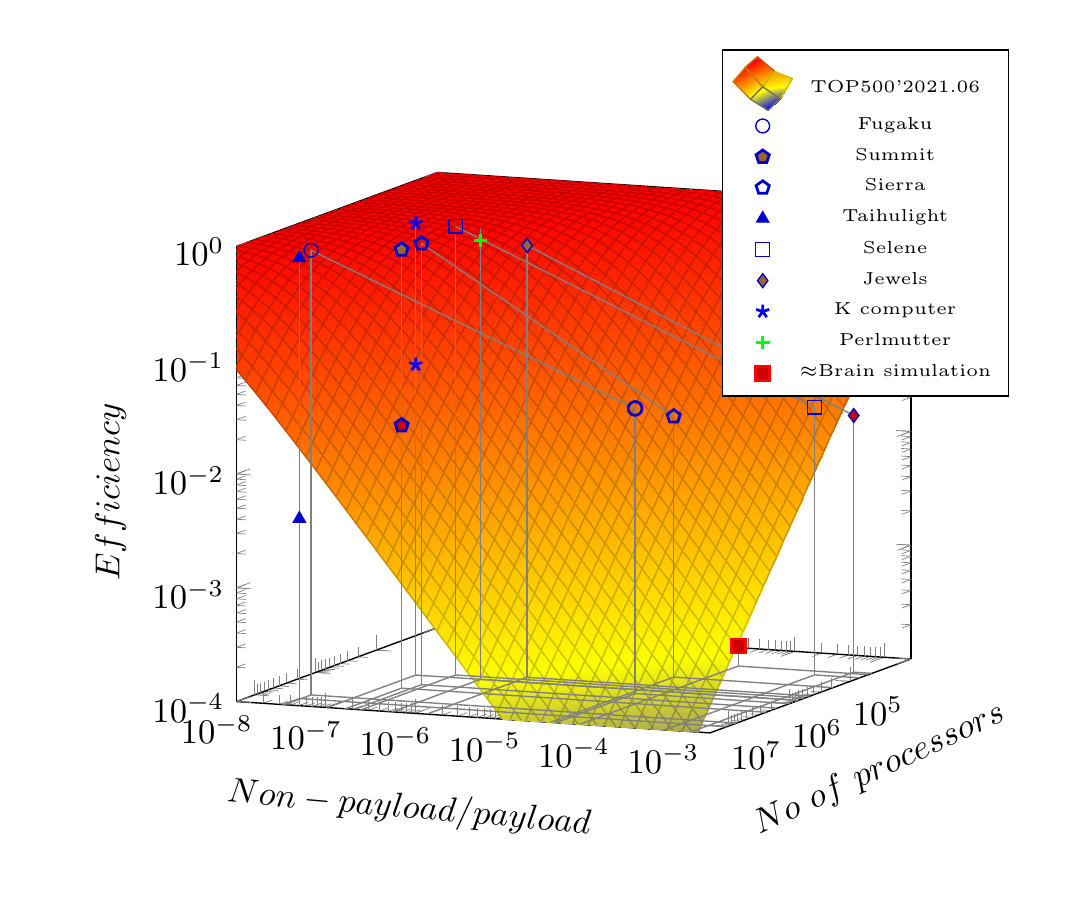}
	\caption{The 2-parameter efficiency surface (in function of parallelization efficiency measured by benchmark \gls{HPL} and  number of processing elements) as concluded from Amdahl's Law (see Eq.~(\ref{eq:soverk})), in first order approximation. Some sample efficiency values for
		some selected supercomputers are shown, measured with benchmarks \gls{HPL} and \gls{HPCG}, respectively.      Also, the estimated efficacy of brain simulation using conventional computing is shown.    \label{fig:EffDependence2020Log}
	}
\end{figure*}

As seen, \textit{the efficiency of a parallelized system is a two-parameter function}
(the corresponding parametrical surface is shown in Fig.~\ref{fig:EffDependence2020Log}), demonstratively underpinning that "\textit{This decay in performance is not a fault of the architecture, but is dictated by the limited parallelism}"~\cite{ScalingParallel:1993}.
Furthermore, that its dependence can be correctly described 
by the properly interpreted Amdahl's Law,
rather than being an unexplained "empirical efficiency".

\subsection{Gustafson's Law\label{sec:GustafsonLaw}}

Partly because of the outstanding achievements of parallelization technology, partly because of issues around practical utilization of Amdahl's Law, a 'weak scaling' (also called Gustafson's Law~\cite{Gustafson:1988}) was also introduced.  Its assumption is that \textit{the computing resources grow proportionally with the task size}, and the speedup (using our notations) is formulated as

\begin{equation}
S = (1-\alpha) + \alpha \cdot N \label{Equ:Gustafson}
\end{equation}

Similarly to Amdahl's Law, the efficiency can be derived for Gustafson's Law as (compare to Eq.~(\ref{eq:soverk}))
\begin{equation}
E(N,\alpha) = \frac{S}{N} = \alpha  + \frac{(1-\alpha)}{N} \label{Equ:GustafsonEff}
\end{equation}

\noindent From these equations immediately follows that speedup (aka parallelization gain) \textit{increases} linearly with the number of processors, \textit{without limitation}; a conclusion that was launched amid much fanfare. They imply, however, some more immediate findings, such as

\begin{itemize}
	\item the efficiency slightly \textit{increases} with the number of processors $N$ (the more processors, the better efficacy) 
	
	\item the non-parallelizable portion of the task either shrinks as the number of processors grows, or \textit{despite that it is non-parallelizable, the portion $(1-\alpha)$ is distributed between the $N$ processors} 
	\item \textit{executing the extra machine instructions needed to organize the joint work need no time}
	\item  \textit{all non-payload computing contributions such as
		communication (including network transfer), synchronization, input/output and memory access take no time}
\end{itemize}
However, an error was made in deriving Eq.~(\ref{Equ:Gustafson}): \textit{the $N-1$ processors
	are idle waiting (see the term with subscript '$idle$' below) while the first one is executing the 
	sequential-only portion}. Because of this, the \textit{time} 
that serves as the base for calculating\footnote{For the meaning of the terms, the wording "is the amount of time spent (by a serial processor)" is used by the author in~\cite{Gustafson:1988}} the \textit{speed}up
in the case of using $N$ processors is% (see also Fig.~\ref{fig:RelativisticParallel} and related text)

\begin{align*}
T_{N} & =(1- \alpha)_{processing} + \alpha\cdot N + (1- \alpha)\cdot(N-1)_{idle}\\
& = (1- \alpha) \cdot N + \alpha\cdot N \\
& = N\\
\label{Equ:GustafsonTE}
\end{align*}

That is, before fixing the arithmetic error, strange conclusions follow, after fixing it,
the conceptual mistake comes to light: \textit{'weak scaling' assumes that single-processor efficiency can be transferred to parallelized sequential subsystems without loss}. Weak scaling assumes that the efficacy of a system comprising $N$ single-thread processors remains the same as that of a single-thread processor. This  assumption strongly contradicts the experienced 'empirical efficiency' (several hundred-fold deviation from its predicted value) of parallelized systems,
not mentioning the 'different efficiencies'~\cite{VeghHowMany:2020}, see also Fig.~\ref{fig:EffDependence2020Log}.

However,  that "\textit{in practice, for several applications, the fraction of the serial part happens to be very, very small thus leading to near-linear speedups}"\cite{UsesAbusesAmdahl:2001}, mislead the researchers.
Gustafson concluded his "scaling" for several hundred processors only.
The interplay of improving parallelization and general \gls{HW} development (including the non-determinism of modern \gls{HW}~\cite{PerformanceCounter2013}), covered for decades
that this scaling was used far outside of its range of validity.

That is,  Gustafson's Law is simply a misinterpretation of its argument  $\alpha$: a simple function form transforms  Gustafson' Law to Amdahl's Law~\cite{AmdalVsGustafson96}. After making that transformation, the two (apparently very different) laws become identical. However, as suspected by~\cite{AmdalVsGustafson96}: "\textit{Gustafson's formulation
	% (also called 'strong scaling') 
	gives an illusion that as if N can increase indefinitely}".
Although collective experience showed, that it was not valid for the case of systems comprising an ever higher number of processors (an "empirical efficiency" appeared),
and later researchers measured "two different efficiencies"~\cite{DifferentBenchmarks:2017} for the same supercomputer (under different workloads),
the "weak scaling" was not suspected to be responsible for the issues.

\textit{"Weak scaling" omits all non-payload (but needed for the operation) activities, such as interconnection time,
	physical size (signal propagation time), accessing data in an amount exceeding cache size, synchronization of different kinds}, that are undoubtedly present when working with \gls{ANN}s.
In other words, \textit{'weak scaling" neglects the temporal behavior},
a crucial science-based feature of computing~\textbf{\cite{VeghTemporal:2020}}.
This illusion led to the moon-shot of targeting to build
supercomputers with computing performance well above feasible (and reasonable) size~\textbf{\cite{VeghHowMany:2020}} and leads to false conclusions in the case of clouds.
Because of this, \textit{except some very few neuron systems, "weak scaling" cannot be safely used for \gls{ANN}s, even as a rough approximation}.

\subsection{Time-aware scaling\label{sec:ModernLaw}}

The role of $\alpha$ was theoretically
established~\cite{Karp:parallelperformance1990},
and the phenomenon itself,
that the efficiency (in contrast with Eq.~(\ref{Equ:GustafsonEff})) \textit{decreases} as the
number of processing units \textit{increases},
is known since decades~\cite{ScalingParallel:1993} (although it was not formulated in the functional form given by Eq.~(\ref{eq:soverk})).
%In the past decades, however, the theory was somewhat faded mainly due to quick development of parallelization technology and increase of single-processor performance;
%and finally, because 'weak scaling' was used
%to calculate the expected performance values, 
%in many cases outside of its range of validity. 
The 'gold rush' for building exascale computers
made finally evident, that under the extreme
conditions represented by the need of millions of processors,
'weak scaling' leads to false conclusions. It had to be admitted that it  "\textit{can be seen in our current situation where the historical ten-year cadence between the attainment of megaflops, teraflops, and petaflops has not been the case for exaflops}"~\cite{ExascaleGrandfatherHPC:2019}.
It looks like, however, that in feasibility studies of supercomputing using
parallelized sequential systems,  and an analysis,
whether building computers of such size is feasible (and reasonable) remained (and remains) out of sight either in USA~\cite{NSA_DOE_HPC_Report_2016,Scienceexascale:2018} or in  EU~\cite{EUActionPlan:2016}
or in Japan~\cite{JapanExascale:2018} or in China~\cite{ChinaExascale:2018}.

Fig.~\ref{fig:EffDependence2020Log}
depicts the two-parameter efficiency surface stemming from the time-aware interpretation of Amdahl's law. 
On the parametric surface, described by Eq.~(\ref{eq:soverk}), some measured efficiencies of present top supercomputers are also depicted, only to illustrate some general rules. 
The \gls{HPL}\footnote{http://www.netlib.org/benchmark/hpl/} efficiencies are sitting on the surface, while
the corresponding \gls{HPCG}\footnote{https://www.epcc.ed.ac.uk/blog/2015/07/30/hpcg} values are much below those values.
The conclusion drawn here was that "\textit{the supercomputers have two different efficiencies}"~\cite{DifferentBenchmarks:2017},
because that experience cannot be explained in the frame of 
'classic computing paradigm' and/or 'weak scaling'.

%As Figure~\ref{fig:EffDependence2018LogA} witnesses,
Supercomputers $Taihulight$, $Fugaku$ and $K~computer$ stand out from the "millions core" middle group. Thanks to its 0.3M cores, $K~computer$ has the best efficiency for $HPCG$ benchmark, while $Taihulight$ with its 10M cores the worst one.   
The middle group follows the rules~\textbf{\cite{VeghHowMany:2020}}. For $HPL$ benchmark: the more cores, the lower efficiency.
It looks like the community experienced the effect of the two-dimensional efficiency.
The top supercomputers run \gls{HPL} benchmark with using all their cores,
but some of them only use a fragment of their cores to measure performance with \gls{HPCG}. This reduction happens because of the inflection point: as can be concluded from the figure, increasing their nominal performance by an order of magnitude,
decreases their efficiency (and so: payload performance) by more than an order of magnitude.
For $HPCG$ benchmark: 
the "roofline"~\cite{WilliamsRoofline:2009} of that communication intensity was already reached,
all computers have about the same efficiency.

Beginning with June 2021, the datum "Measured cores" are not provided any more; covering this aspect.
The presumable reason of this new trend is that in this way the measures \gls{HPCG}/\gls{HPL} ratio gets much higher, providing the illusion that the vast supercomputers became more suitable for real-life tasks.	
For more discussion on supercomputers, see~\textbf{\cite{VeghHowMany:2020}} (and its continuous upgrades or arXiv).

\section{Performance limit of processor-based AI systems}\label{fig:AIperformanceLimit}

\subsection{General considerations}
As discussed in~\textbf{\cite{VeghHowMany:2020}}, payload performance $P(N,\alpha)$
of parallelized systems comprising $N$ processors  is
described\footnote{At least in a first approximation, see~\textbf{\cite{VeghHowMany:2020}}} as

\begin{equation}
P(N,\alpha) = \frac{N\cdot P_{single}}{{N\cdot \left(1-\alpha\right)+\alpha}}\label{eq:Ppayload}
\end{equation}

\noindent where $P_{single}$ is the single-thread performance of individual processors, and $\alpha$ is describing
parallelization of the given system for the given workload (i.e., it depends on both of them). 

\noindent This simple formula explains why \textit{payload performance  of a system
	is not a linear function of its nominal performance} and why
in the case of very good parallelization ($(1-\alpha)\ll1$)
and low $N$, this non-linearity cannot be noticed.
In contrast with the prediction of 'weak scaling', \textit{payload performance} and \textit{nominal performance} differ by a factor, growing with the number of cores.
This conclusion is well-known, but forgotten: "\textit{This decay in performance is not a fault of the architecture, but is dictated by the limited parallelism}"~\cite{ScalingParallel:1993}.

The key issue is, however, that one can hardly calculate the value of $\alpha$ for the present
complex \gls{HW}/\gls{SW} systems from their technical data,
although some estimated values can be derived.
For supercomputers, however, one can derive a theoretical "best possible", and already achieved "worst case" values~\textbf{\cite{VeghHowMany:2020}}. It gives us reasonable confidence that those values deviate only within a factor of two. We cannot expect similar results for \gls{ANN}s. There are no generally accepted benchmark computations, and also there are no standard architectures\footnote{Notice, that selecting a benchmark also directs the architectural development: the benchmarks \gls{HPL} and \gls{HPCG} result in different rankings.}.
Using a benchmark means a particular workload, and comparing the results of even a standardized  \gls{ANN} benchmark on different architectures is conceptually as little useful as comparing the results of benchmarks \gls{HPL} and \gls{HPCG} on the same architecture.

Recall also, that at a large number of processors, the
internal latency of processor also matters.
Following the failure of supercomputer Aurora'18, Intel admitted: "\emph{\textit{Knights Hill} was canceled and instead be replaced by a "\emph{new platform and new microarchitecture specifically designed for exascale}"}"~\cite{IntelDumpsXeonPhi:2017}. We expect that shortly it shall be admitted that building large-scale \gls{AI} systems is simply not possible based on the old architectural principles~\cite{DeepNeuralNetworkTraining:2016}.
The potential new architectures, however, require a new computing paradigm (considering both temporal behavior of computing systems~\textbf{\cite{VeghTemporal:2020}}, and the old truth that "more is different"~\cite{MoreIsDifferent1972}), that can give a proper reply to power consumption and performance issues of -- among others,  \gls{ANN} -- computing.

\subsection{Communication-to-computation ratio}
As we learned decades ago, "\textit{the inherent communication-to-computation ratio in a
	parallel application is one of the important determinants
	of its performance on any architecture}"~\cite{ScalingParallel:1993},
suggesting that communication can be a dominant contribution to system’s non-payload performance.
In the case of neural simulation, a very intensive communication must take place, so
the non-payload to payload ratio has a significant impact on the performance of
\gls{ANN}-type computations. That ratio and the corresponding workload type are closely related: using a specific benchmark implies using a specific communication-to-computation ratio. In the case of supercomputing, the same workload is running on (nearly) the same type of architecture, which is not the case for \gls{ANN}s.
Communication is implemented through \gls{I/O} instructions;
this is why "\textit{artificial intelligence, \dots it's the most disruptive workload from an I/O pattern perspective}"\footnote{ https://www.nextplatform.com/2019/10/30/cray-revamps-clusterstor-for-the-exascale-era/}

\subsection{Computing benchmarks}

There are two commonly used benchmarks in supercomputing.
Historically, benchmark \gls{HPL} was used to compare different configurations by executing a standard computing task.
However, their performance depends very much on their workload type.
Although \gls{HPL} is excellent for racing purposes (produces high figures), its behavior on vast configurations strongly deviates from that of "real-life" tasks: the surface in Fig.~\ref{fig:EffDependence2020Log} represents a kind of theoretical upper limit for distributed processing.

The \gls{HPL} class tasks essentially need communication only at
the very beginning and at the very end of the job. Real-life programs, however, usually work in a \replaced{different}{non-standard} way. Because of this reason,  a couple of years ago, the community introduced benchmark \gls{HPCG}: the collective experience shows that the payload performance is much more accurately approximated by \gls{HPCG} than by \gls{HPL}, because real-life tasks need much more communication than  \gls{HPL}.
Importantly, since their interconnection quality improved considerably, supercomputers show different efficiencies when using various benchmark programs~\cite{DifferentBenchmarks:2017}. Their efficiencies differ by a factor of ca. 200-500 (a fact that remains unexplained in the frame of "weak scaling"), when measured by \gls{HPL} and \gls{HPCG}, respectively.

\begin{figure}
	\maxsizebox{\columnwidth}{!}
	{
		\includegraphics[width=\textwidth]{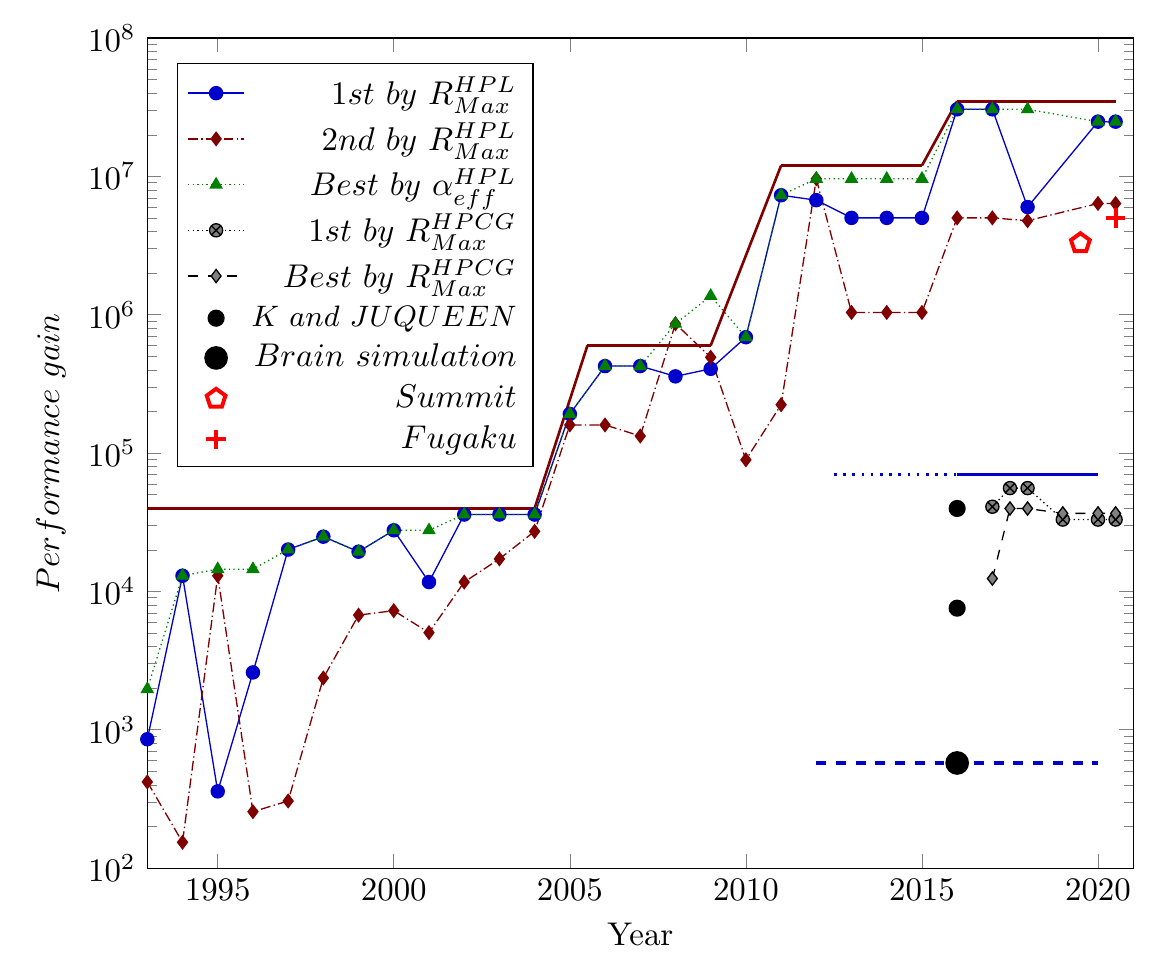}
	}
	\vspace{-\baselineskip}
	\caption{{Performance gain of supercomputers in function of their year of construction, under different workloads. The diagram lines
			display the measured values
			derived using \gls{HPL} and \gls{HPCG} benchmarks, for the TOP3 supercomputers in the gives years.
			The small black dots mark the performance data of
			supercomputers $JUQUEEN$ and $K$ as of 2014 June,
			for \gls{HPL} and \gls{HPCG} benchmarks, respectively.
			The big black dot denotes the payload performance of the system used by \cite{NeuralNetworkPerformance:2018}.
			The saturation effect can be observed for both
			\gls{HPL} and \gls{HPCG} benchmarks.
			\label{fig:RooflineBrain}}
	}
\end{figure}

In the \gls{HPL} class,  the communication intensity is the lowest possible one:
computing units receive their task (and parameters) at the beginning
of computation, and they return their result at the very end.
The core orchestrating their work must deal with the fellow cores only in these periods, so the communication intensity is proportional to the number of cores in the system. Notice the need to queue requests at the beginning and the end of the task.

In the \gls{HPCG} class, iteration occurs:  the fellow cores return the result of one iteration to the coordinator core, which makes sequential operations: receives and re-sends the parameters, then it needs to compute new parameters before sending them back to the fellow cores. The program repeats the process several times. As a consequence, \textit{the non-parallelizable fraction of benchmarking time grows proportionally to the number of iterations}.
The effect of that extra communication decreases the achievable performance roofline~\cite{WilliamsRoofline:2009}: as shown in Fig.~\ref{fig:RooflineBrain}, the \gls{HPCG} roofline is about 200 times smaller than the \gls{HPL} one. As depicted, for benchmark \gls{HPL} multiple rooflines can be located. The highest roofline can be attributed to a processor having slightly different computing principle~\cite{CooperativeComputing2015}. The second highest roofline (dominated by the calculation itself) is the commonly achievable performance gain\footnote{Notice that even using coherence bus or very clever positioning of $L_2$caches cannot help a lot; see the effect of the high number of processors in Fig.~\ref{fig:EffDependence2020Log}.}; their theoretical expectation matches the empirical value~\cite{VeghHowMany:2020}. The third roofline shows the effect of
the improved connection technology. The fourth roofline (dominated mainly by the internal interconnection) shows how much performance can can be achieved without using expensive "racing" parallelization technologies. For benchmark \gls{HPCG}, only one roofline exists: the calculation dominates. Neither direct interconnection of processors nor advanced interconnection technology can increase the performance gain. \added{As discussed below, for another types of calculations (non-standard benchmarks), the roofline can be even lower: \cite{AngeloLearningBioinformatics:2014} measured a value as low as 30
	for a certain kind of applications.
} The "roofline for brain simulation" is guessed from one single measurement datum. \added{That datum was measured using a only a small fragment (about 1\%) of available cores (see section~\ref{sec:QuantalTime}), so the real efficiency (see also section~\ref{sec:ModernLaw} on benchmarking supercomputers using only a fragment of their available cores) can be about two orders of magnitude lower, i.e., in the order of a dozen.}

\begin{figure*}
	\includegraphics[width=.9\textwidth]{%fig/
		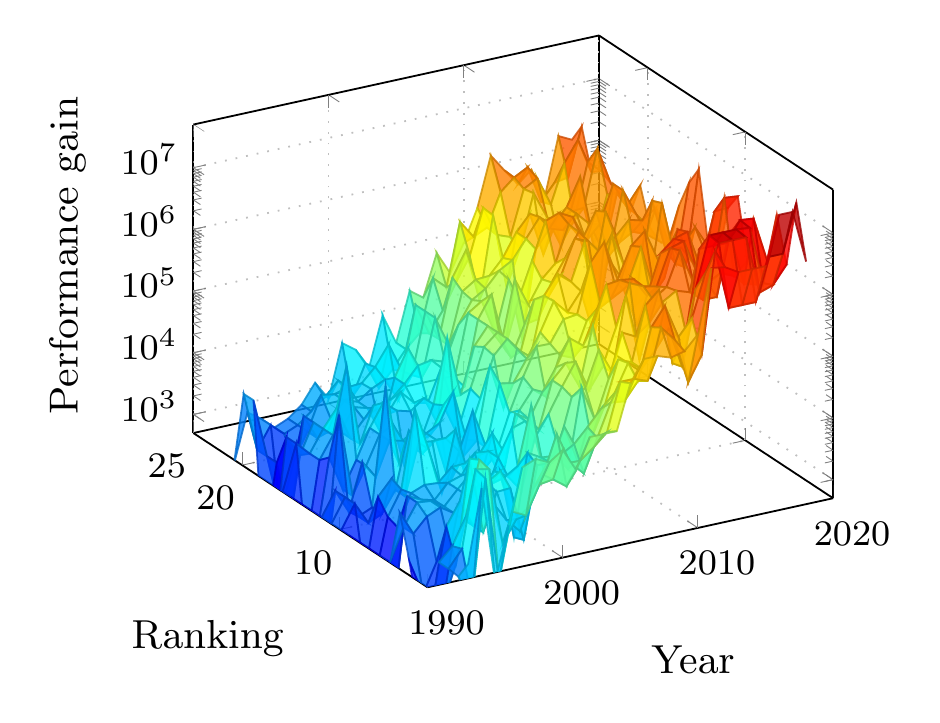}
	\vspace{-.5cm}
	\caption{History of supercomputing in terms of performance gain:  performance values of the first
		25 supercomputers, in function of year of their ranking. Data measured using benchmark \gls{HPL} 
		\label{fig:SupercomputerGainHillside}
	}
	\vspace{-\baselineskip}
\end{figure*}

Unfortunately, an \gls{ANN} workload, in general, cannot be defined, so one cannot run one single benchmark
to guess the performance characteristics of a  new application.
The performance gain for \gls{ANN}s can be guessed to be around or above that of the brain simulation\footnote{\added{Our reasoned guess is in good accordance with the experimental evidence~\cite{DeepNeuralNetworkTraining:2016}:"\textit{strong scaling is stalling after only a few dozen nodes}"}}; depending on \gls{ANN}'s architecture. The neural communication, using a vast number of simultaneous communication, combined with 
the idea of using a single high speed bus, introduces an additional bottleneck: a very different type of workload, see section~\ref{sec:bus}.
Because of this different workload, the above benchmarks cannot be used to estimate execution time of \gls{ANN}-type tasks.

As expressed by Eq.~(\ref{eq:Ppayload}),
the resulting performance of parallelized computing systems depends on both single-processor performance and performance gain.
To separate these two factors, 
Fig.~\ref{fig:SupercomputerGainHillside} displays the \textit{performance gain} of supercomputers in the function of their year of construction and ranking in the given year.
Two 'plateaus' can be localized before the year 2000 and after the year 2010 also, unfortunately; underpinning Amdahl's Law and refuting Gustafson's Law, and also confirming the prediction "\textit{Why we need Exascale and why we won't get there by 2020}"\cite{WhyNotExascale:2014}.
The "hillside" reflects the enormous development of interconnection technology between the years 2000 and 2010 (for more details see~\textbf{\cite{VeghHowMany:2020}}). For the reason of the "humps" around the beginning of the second plateau, see section~\ref{sec:accelerators}.
Unfortunately, different individual factors (such as interconnection quality, using accelerators and clustering, using on-chip memories, or using slightly different computing paradigm, etc.) cannot be separated in this way. However, some limited validity conclusions can be drawn.

\begin{figure*}
	\includegraphics[width=\textwidth]{%fig/
		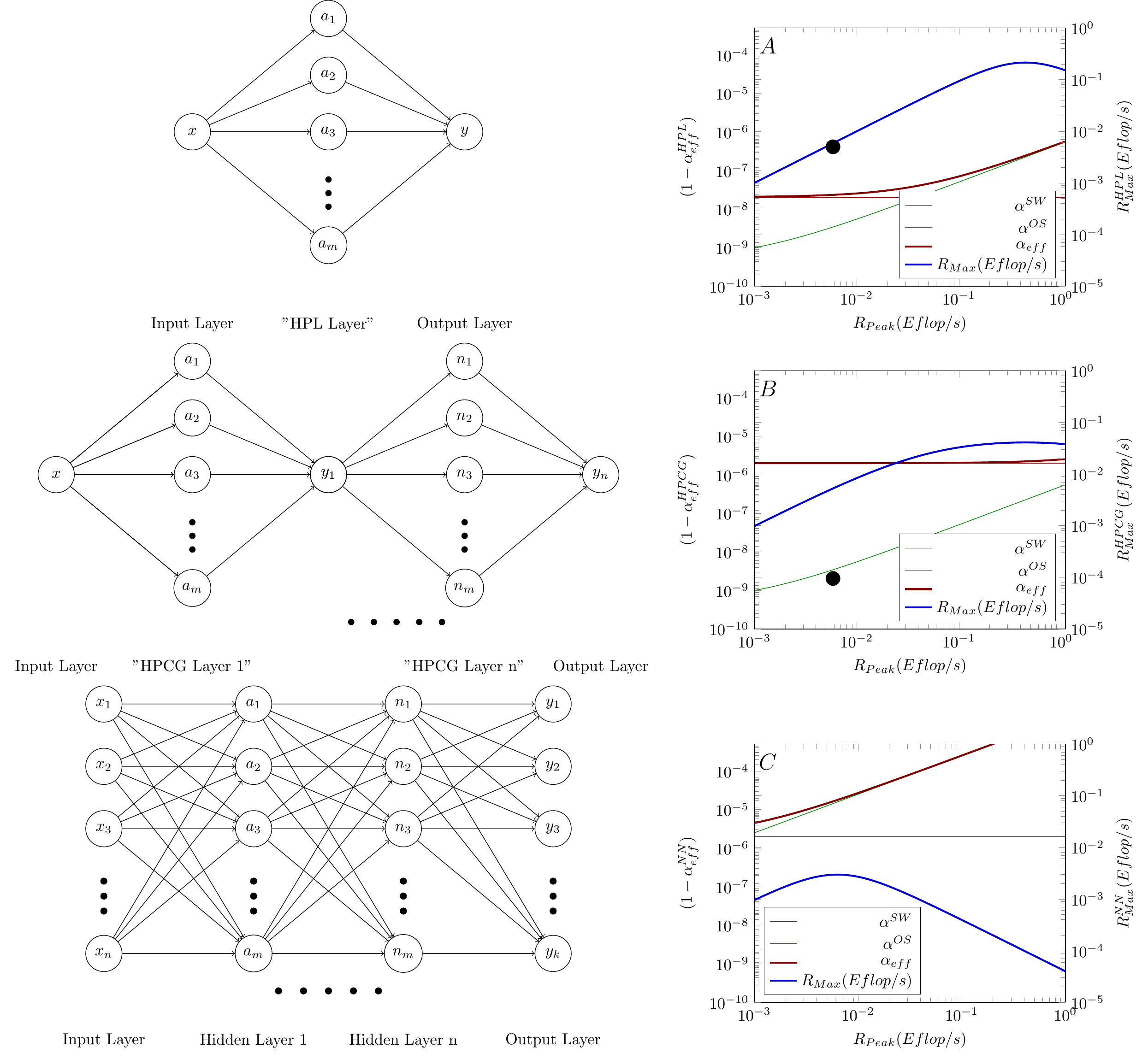}
	\vspace{-\baselineskip}
	\caption{Different communication/computation intensities of the applications
		lead to different payload performance values in the same supercomputer system. Left column: models of computing intensities for different benchmarks. Right column: the corresponding payload performances and $\alpha$ contributions in function of the nominal performance of a fictive supercomputer ($P=1Gflop/s$ @ $1GHz$).
		The blue diagram lines refer to the right hand scale ($R_{Max}$ values), all others ($(1-\alpha_{eff}^{X})$ contributions) to the left hand scale. The figure is purely illustrating the concepts; the displayed numbers are somewhat similar to the real ones. 		\label{fig:AlphaContribBenchmark}
	}
	\vspace{-\baselineskip}
\end{figure*}

\subsection{Workload type}

The role of the workload came to light after that 
interconnection technology was greatly improved, and as a
consequence, the \textit{benchmarking computation} (defining
\textit{the type of workload}) became the dominating contributor,
defining value of $\alpha$ (and as a consequence, payload performance), for a discussion see~\textbf{\cite{VeghHowMany:2020}}.
The overly complex Fig.~\ref{fig:AlphaContribBenchmark}
illustrates the  phenomenon, why and how the payload performance of a
configuration depends on the application it runs. Notice that at low nominal performance values,
the payload performance depends linearly on the nominal performance (the blue diagram line), and only slightly depends on the workload type.

The performance breakdown shown in Fig.~\ref{fig:AlphaContribBenchmark} was experimentally measured by~\protect{\cite{ScalingParallel:1993,AngeloLearningBioinformatics:2014}}, ~\protect{\cite{NeuralScaling2017}}(Fig.~7)
	and~\protect{\cite{NeuralNetworkPerformance:2018}}(Fig.~8), but it used not to be a subject of studies.
The fact that speedup diagram line turns back at a critical number of processor, was noticed early~\cite{ScalingParallel:1993}.
	After exceeding a critical number of cores, housekeeping gradually becomes the dominating factor of the performance limitation, and leads to a decrease in the payload performance: "\textit{there comes a point when using more \gls{PU}s \dots actually increases the execution time rather than reducing it}".  In that paper (at a
		different workload and architecture) the achievable parallelization gain was about 8, and it was achieved using
		20-30 processors. Recently, there are few experimental investigations in this direction.
		One of the rare exceptions is~\cite{AngeloLearningBioinformatics:2014}. The careful systematic investigation of results of  running bioinformatics applications
		pointed out that the speedup curve has a maximum, and breaks down for a higher number of processor cores:  
		"\textit{The execution time and the speed-up on IPDATA reach the best values within about 90 processors}. Furthermore, that \dots 
		"\textit{the parallel version is up to 30 times faster than the serial one}". \added{For \gls{ANN}s, it is just a few dozens~\cite{DeepNeuralNetworkTraining:2016} where "strong scaling" stalls.} For different applications (workloads) these figures are of course different, but the conclusion persists.

Fig.~\ref{fig:AlphaContribBenchmark} compares three workloads (owing
different communication intensity).
In the top and middle figures, the communication
intensities of the standard supercomputer benchmarks $HPL$ and $HPCG$ are displayed in the style of AI networks.
The "input layer" and "output layer" are the same,
and comprise the initiating node only, while the
other "layers" are again the same: the rest of the cores.
Subfigure~\ref{fig:AlphaContribBenchmark}.C depicts an AI network comprising
$n$ input nodes and $k$ output nodes, furthermore the
$h$ hidden layers are comprising $m$ nodes. The communication-to-computation intensity~\cite{ScalingParallel:1993} is, of course, not proportional in the cases of subfigures, but the figure illustrates excellently how the communication need of different computer tasks changes with the type of the workload.

As can be easily seen from the figure,
in the case of benchmark $HPL$, the initiating node must
issue $m$ communication messages and collect $m$ returned results,
i.e., the execution time is $O(2m)$.
In the case of benchmark $HPCG$ this execution time is
$O(2Nm)$ where $N$ is the number of iterations
(one cannot directly compare the execution times because of the different amounts of computations and the different amounts of sequential-only computations).

Fig.~\ref{fig:AlphaContribBenchmark}.A displays the case of minimum communication, and Fig.~\ref{fig:AlphaContribBenchmark}.B a moderately increased one
(corresponding to real-life supercomputer tasks).
As nominal performance increases linearly and payload performance decreases inversely with the number of cores,
at some critical value where an inflection point occurs, the
resulting performance starts to fall.
The resulting  non-parallelizable fraction sharply decreases efficacy
(in other words: performance gain or speedup) of the system~\cite{VeghPerformanceWall:2019}.
This effect was noticed early~\cite{ScalingParallel:1993}, under different technical conditions, but somewhat faded due to development of parallelization technology.

The non-parallelizable fraction (denoted in the figure by $\alpha_{eff}^{X}$) of a computing task comprises components $X$ of different origin. As already discussed, and was noticed decades ago, "\textit{the inherent communication-to-computation ratio in a
	parallel application is one of the important determinants
	of its performance on any architecture}"~\cite{ScalingParallel:1993},
suggesting that \textit{communication can be a dominant contribution to the system’s performance}.

The workload in \gls{ANN} systems comprises components
of type "computation" and "communication" (this time also
involving data access and synchronization, i.e., everything that is 'non-computation'). As logical interdependence between those contributions
is strictly defined, payload performance of the system is limited
by both factors, and the same system (maybe even within the same workload, case by case) can be either computing-bound and communication bound, or both.

Notice that supercomputers showing the breakdown depicted in Fig.~\ref{fig:AlphaContribBenchmark}, are not included
in history depicted in Fig.\ref{fig:SupercomputerGainHillside}. Aurora'18 failed, Aurora'21 semi-failed, Gyokou was withdrawn, "Chinese decision-makers decided to withhold the country’s newest Shuguang supercomputers even though they operate more than 50 percent faster than the best current US machines"\footnote{https://www.scmp.com/tech/policy/article/3015997/china-has-decided-not-fan-flames-super-computing-rivalry-amid-us}.
Also, $Fugaku$ stalled~\cite{DongarraFugakuSystem:2020} at some 40\%
of its planned capacity.

Notice that a similar hillside cannot be drawn for benchmark \gls{HPCG}, because of two reasons.
On one side, \gls{HPCG} measurement started only a few years ago. On the other side, top supercomputers publish data measured with cores less than the number of cores used to measure \gls{HPL}. 
Recall that efficiency is a two-parameter function (see Fig.~\ref{fig:EffDependence2020Log}).
 For "real-life" programs, represented by \gls{HPCG},
this critical number is much lower than in the case of \gls{HPL}.
There is a real competition between the different contributions to dominate system's performance. As   demonstrated in Fig.~6 in~\textbf{\cite{VeghHowMany:2020}}, before 2010, running  both benchmarks \gls{HPL} and \gls{HPCG}
on a top supercomputer was a \textit{communication-bound }task, 
since 2010 \gls{HPL} is a computing-bound task, while \gls{HPCG} persisted to be a \textit{communication-bound} task.  This is why some supercomputers provide their \gls{HPCG} efficiency  measured only with a fragment of their cores: the \gls{HPCG} 
"roofline" is reached at that lower number of cores. Adding more cores does not increase their payload performance, but decreases their efficiency.    

%\subsubsection{The operand length}
\subsection{Accelerators}\label{sec:accelerators}

%The non-computational (and other non-payload) times, by their definition, are non-parallelizable. Because of this, they contribute to the
%sequential-only portion of the task, and so they make the efficiency of the
%large systems drastically worse. 

As a side-effect of "weak scaling", it is usually presumed that decreasing the time needed for the payload contribution affects the efficiency of \gls{ANN} systems linearly. However, it is not so.
As discussed in detail in~\textbf{\cite{VeghDoWeKnow:2020}}, we also change the non-payload to payload ratio that defines the system's efficiency. We mention two prominent examples here: using shorter operands (move less data
and perform less bit manipulations) and to mimic the operation
of a neuron in an entirely different way: using quick analog
signal processing rather than slow digital calculation.

It is a common fallacy that benchmark \textit{HPL-AI}
is benchmark for \gls{AI} systems. Actually, it means "\textit{The High Performance LINPACK for Accelerator Introspection}" (HPL-AI), and that benchmark seeks to highlight the convergence of HPC and artificial intelligence (AI) workloads.\footnote{https://www.icl.utk.edu/hpl-ai/} It has not much to do with \gls{AI}, except that it uses the operand length common in \gls{AI} tasks. \gls{HPL}, similarly to \gls{AI}, is a \textit{workload type}\footnote{Even 
	\textit{https://www.top500.org/lists/top500/2020/06/} mismatches \textit{operand length} and \textit{workload}: "In single or further reduced precision, which are often used in machine learning and AI applications, Fugaku's peak performance is over 1,000 petaflops (1 exaflops)". }.
Researchers  succeeded to achieve more than three times better
performance gain (3.01 for $Summit$ and 3.42 for $Fugaku$), that (as correctly stated in the announcement)
"\textit{Achieving a 445 petaflops mixed-precision result on HPL (equivalent to our 148.6 petaflops DP result)}"~\cite{MixedPrecisionHPL:2018}, i.e. the peak DP performance did not change. \textit{However, this naming convention suggests the illusion that when using supercomputers for \gls{AI} tasks and using half-precision, one can expect this payload performance.}

Unfortunately, \textit{this achievement comes from accessing less data in memory and using quicker operations on shorter operands
	rather than reducing communication intensity}.
For \gls{AI} applications, limitations remain the same
as described above;
except that when using Mixed Precision,
the power efficiency shall be better by a factor of nearly four, compared to the power efficiency measured using double precision operands.\footnote{
	Similarly, exchanging data directly between processing units~\cite{CooperativeComputing2015} (without using the global memory)
	also enhances $\alpha$ (and payload performance)~\cite{TaihulightHPCG:2018},
	but it represents a (slightly) different computing paradigm.}

We expect that when using half-precision (FP16) rather than double precision
(FP64) operands in the calculations, four times less
data are transferred and manipulated by the system.
The measured power consumption data underpin the statement.
However, the system's computing performance is only slightly more than three times higher than using 64-bit (FP64) operands.
The non-linearity has its effect even in this simple case (recall that \gls{HPL} uses minimum communication). In the benchmark, the housekeeping activity (data access, indexing, counting, addressing) also takes time. Concerning the temporal behaviour~\textbf{\cite{VeghTemporal:2020}} of the operation, in this case, the data transmission time $T_t$ is the same, the data processing time (due to the shorter length of operands) changes, and so the apparent speed changes non-linearly.
Even, the
measured performance data enabled us to estimate execution time with zero precision\footnote{Without dedicated measurement, no more accurate estimations are possible} (FP0) operands, see~\textbf{\cite{VeghHowMany:2020}}.

Another plausible assumption is that if we use quick analog signal processing
to replace the slow digital calculation, as proposed in~\cite{RecipeMemristor:2020,NatureBuildingBrain:2020,ContinuousTimeCrossbarMemristor:2021},
our computing system gets proportionally quicker.
Presumably, on systems comprising just a few neurons, one can measure a considerable, but less than expected, speedup. 
The housekeeping becomes more significant than in the case of purely digital processing. In a hypothetical measurement, the speedup would be much less than the ratio of the corresponding analog/digital processing times, even in the case of \gls{HPL} benchmark. Recall that here the workload is of \gls{AI} type, with much worse parallelization (and non-linearity).
As a consequence, one cannot expect a considerable speedup in large neuromorphic systems.
For a detailed discussion of introducing new effect/technologies/materials, see~\textbf{\cite{VeghTemporal:2020}}.

\added[comment=Citing recent publication]{It sounds good that "\textit{The analog memristor array is effectively the neural
	network laid out in the form of a crossbar, which can perform the
	entire operation \textbf{in one clock cycle}}"~\cite{BrainInspiredBlocks:2020}. In brackets, however fairly added, that \textit{"(not counting the clock cycles that
	may be required to fetch and store the input and output data)"}.
Yes, all operands of the memristor array must be transferred to its input section (and previously, they must be computed or otherwise produced),
furthermore the results must be transferred from its output section to their destination. The effective computing time of the memristor-related operations shall be compared to conventional operations' effective time,
from the beginning to the end of the computing operation,
to make a fair comparison. (The problem persists even if 
continuous-time data representation~\cite{ContinuousTimeCrossbarMemristor:2021}
is used.)

The temporal behavior of components and their materials can easily be misidentified in the time-unaware model. Five decades ago, even
	\textit{memristance} has been introduced~\cite{MissingMemristor:2008} as a fundamental electrical component, meaning that the memristor's electrical resistance is not constant but depends on the history of current that had previously flowed through the device.
	There are, however, some serious doubts as to whether a genuine memristor can actually exist in physical reality~\cite{RejectingMemristor:2018}.
	In the light of our analysis, some temporal behavior definitely exists; the question is how much it is related to material or biological  features, if our time-aware computing method is followed.
}

Besides, adding analog components to a digital processor has its price. Given that a digital processor cannot handle resources outside of its world, one must call the \gls{OS} for help. That help, however, is expensive in terms of execution time. The required context switching takes time in the order of executing $10^4$ instructions~\cite{armContextSwitching:2007,Tsafrir:2007}, which dramatically increases the time of housekeeping and the total execution time. This effect makes the systems' non-payload to payload ratio much worse than it was before introducing that enhancement.

\subsection{Timing of activities}
In \gls{ANN}s, the data transfer time
must be considered seriously.  
In both biological and electronic systems, both the distance between
entities of the network, and the signal propagation speed is finite.
Because of this, in physically large-sized and/or intensively communicating systems, the
"idle time" of processors defines the final performance that
a parallelized sequential system can achieve.
In conventional computing systems, 'data dependence' limits achievable parallelism: we must compute data before using it as an argument for another computation.
Although, of course, also in conventional computing, the computed  data must be delivered to the place of their second utilization, thanks to "weak scaling"~\cite{Gustafson:1988}, this "communication time" is neglected.
For example, scaling of matrix operations and "sparsity", mentioned in \cite{BrainInspiredBlocks:2020}, work linearly \textit{only} if 
data transfer time is neglected.

Timing plays an important role in all levels of computing, from gate-level processing to
clouds connected to the Internet. In~\textbf{\cite{VeghTemporal:2020,VeghMissingSecondDraft:2020}}, the example 
describing temporal operation of a one-bit adder provides a nice example,
that although the \textit{line-by-line compiling
	(sequential programming, called also Neumann-style programming~\cite{BackusNeumannProgrammingStyle}),
	formally introduces only logical dependence, through its technical implementation
	it implicitly and inherently introduces a temporal behavior, too}.

In neuromorphic computing, including \gls{ANN}s, the transfer time is a vital part of information processing.
A biological brain must deploy a "speed accelerator" to ensure that the control signals arrive at the target destination before the arrival of the controlled messages, despite that the former derived from a distant part of the brain~\cite{BuzsakiGammaOscillations:2012}.
\textit{This aspect is so vital in biology that the brain deploys many cells with the associated energy investment 
	to keep the communication speed higher for the control signal.}
Computer technology cannot speed up communication selectively, as in biology. It is also impossible to keep part of the system for a lower speed selectively: the propagation speed of electromagnetic waves is predefined. However, as discussed in~\textbf{\cite{VeghDoWeKnow:2020}},
\textit{handling data timing adequately, is vital, especially for bio-mimicking \gls{ANN}s}.

\subsection{The layer structure}
The bottom part of Fig.~\ref{fig:AlphaContribBenchmark} depicts, how \gls{ANN}s
are supposed to operate. The life begins in several input channels (rather than one as in \gls{HPL} and \gls{HPCG} cases), that would be advantageous.
However, the system must communicate its values to \textit{all} nodes in the top hidden layer:
the more input nodes and the more nodes in the hidden layer(s), the many $times$ more communication is required for the operation. The same situation also happens when the first hidden layer communicates data to the second one, except that here \textit{the square of the number of nodes} is to be used as a weight factor of communication.

Initially, $n$ input nodes
issue messages, each one $m$ messages (queuing\#1) to nodes
in the first hidden layer, i.e., altogether $nm$ messages.
If one uses a commonly used shared bus to transfer messages, these  $nm$ messages must be queued (queuing\#2).
Also, every single node in the hidden layer receives (and processes) $m$ input messages (queuing\#3).
Between hidden layers, the same queuing is repeated (maybe several times)
with $mm$ messages, and finally, $km$ messages are sent to the output nodes. During this process, the system queues messages (at least) three times.
Notice that using a single high-speed bus, because of the needed arbitration, drastically increases
the transfer time of the individual messages, furthermore changes their timing, see section~\ref{sec:bus}.

To make a fair comparison with benchmarks $HPL$ and $HPCG$,
let us assume one input and one output node.
In this case, the AI execution time is $O(h\times m^2)$,
provided that the \gls{AI} system has $h$ hidden layers.
(Here we assumed that messaging mechanisms
between different layers are independent.
It is not so if they share a global bus.)

For a numerical example: let us assume that
in supercomputers, 1M cores are used.
In \gls{AI} networks, 1K nodes are present in the hidden layers,
and only one input and output nodes are used.
In that case, all execution times are $O(1M)$
(again, the amount of computation is sharply different,
so the scaling can be compared, but not the execution times).
This communication intensity explains why in Fig.~3 %\ref{fig:rooflines}
the $HPCG$ "roofline" falls
hundreds of times lower than that of the $HPL$:
the increased communication need strongly decreases
the systems's achievable performance gain.

Notice that the number of \textit{computation} operations increases with $m$,
while number of \textit{communication} operations with $m^2$. In other words:
the more nodes in the hidden layers, the higher is their communication intensity (communication-to-computation ratio), and because of this, the lower is the efficiency of the system.
Recall, that since AI nodes perform simple computations, compared to the functionality of supercomputer benchmarks,
their communication-to-computation ratio is much higher,
making their efficacy even worse.
The conclusions are underpinned by experimental research~\cite{DeepNeuralNetworkTraining:2016}:
\textit{\begin{itemize}
		\item "strong scaling is stalling after only a few dozen nodes"
		\item "The
		scalability stalls when the compute times drop below the communication
		times, leaving compute units idle. Hence becoming a communication bound
		problem."
		\item "the network layout
		has a large impact on the crucial communication/compute
		ratio: shallow networks with many neurons per layer \dots scale worse than deep networks with less neurons."
	\end{itemize}
}

\subsection{Using high-speed bus(es)\label{sec:bus}}
As discussed in connection with the reasoning, why the internal temporal ratio between
transporting and processing data has significantly changed~\textbf{\cite{VeghTemporal:2020}}:
Moore's observation is (was) valid for electronic density only, but not valid for 
connecting technology, such as buses. On one side, because of the
smaller physical size and the quicker clock signal, on the other side 
the unchanged $cm$-long bus cable, using a serial bus means spending the
overwhelming majority of  apparent processing time  with arbitration
(see the temporal diagram and case study in~\textbf{\cite{VeghTemporal:2020})},
so using a sequential bus is at least questionable in large-scale systems:
\textit{the transfer time is limited by the needed arbitration (increases with the number of neurons!) rather than by the bus speed}.
\textit{"The idea of using the popular shared bus to implement the communication medium is no longer acceptable, mainly due to its high contention."}~\cite{ReconfigurableAdaptive2016}

\begin{figure*}
	\includegraphics[width=.9\textwidth]{%fig/
		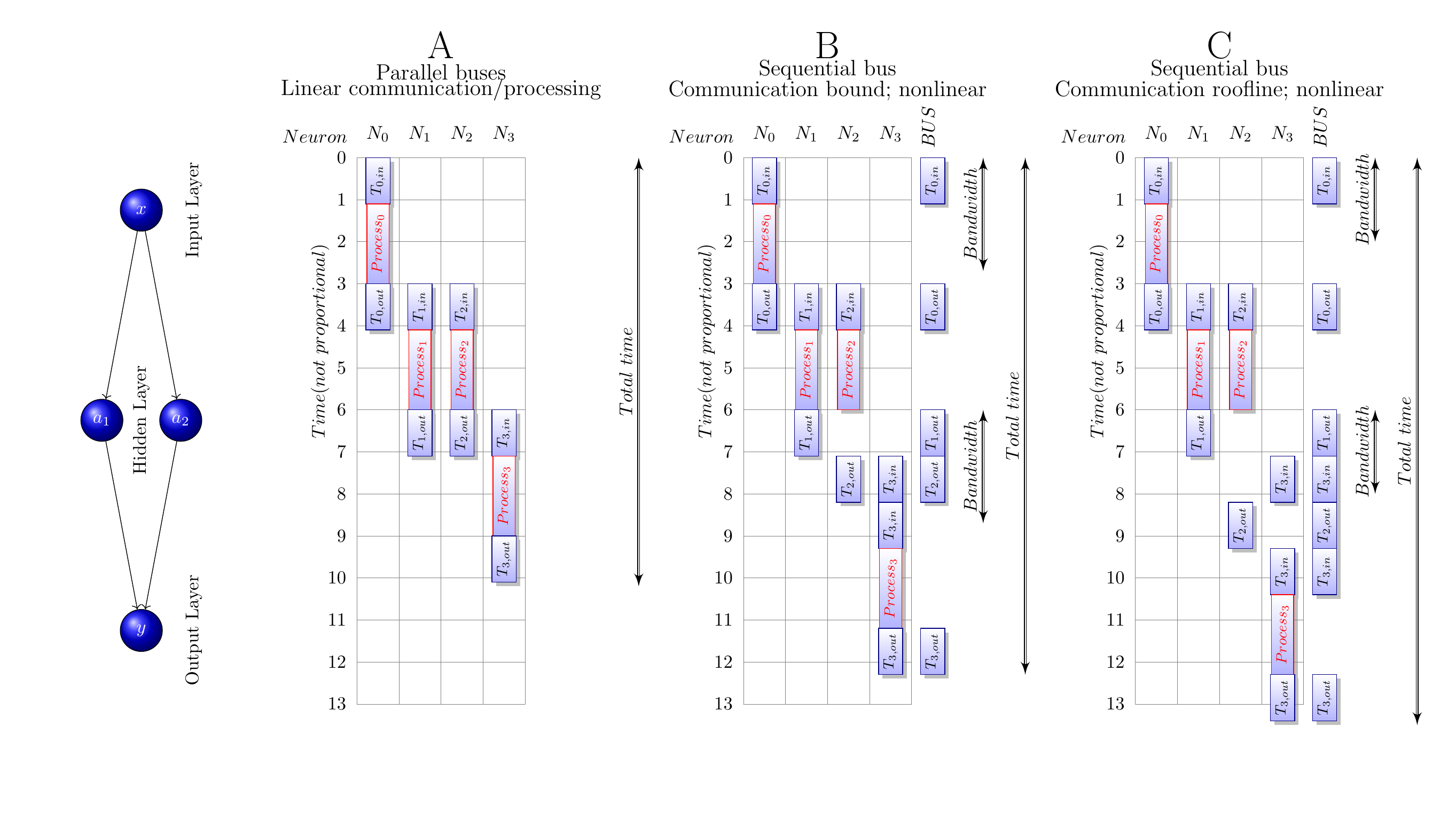}
	\vspace{-.5cm}
	\caption{Implementing neuronal communication in different technical approaches. For legend's details see text. A (the biological implementation): the parallel bus; B and C(the technical implementation): the shared serial bus, before and after reaching the communication "roofline"~\cite{WilliamsRoofline:2009}.\label{fig:Neuronal}
	}
	\vspace{-\baselineskip}
\end{figure*}

The massively "bursty" nature of the data (when imitating biological neural network, the different nodes of the layer want to communicate simultaneously) also makes the case harder.
Communication circuits receive the task of sending data to $N$ other nodes.
What is worse, bus arbitration, addressing and latency, prolong the transfer time (and decrease the system's efficacy).
This type of communicational burst may easily lead to a "communicational collapse"~\cite{CommunicationCollapse:2018},
but it may also produce unintentional "neuronal avalanches"~\cite{NeuronalAvalanches:2003}.
	\added[comment=R1/Q2: Added a sentence for legend's details of Fig.~\ref{fig:Neuronal} ]{}

\textit{The fundamental issue is replacing the private communication channel between biological neurons with the mandatory use of some kind of shared media in technological neurons.} As discussed in~\textbf{\cite{VeghTemporal:2020}},
at a large number of communicating units, sharing the medium becomes the dominant contributor to the time consumption of computing. Its effect can be mitigated using different technological implementations. Still, the conclusion persists: its technical implementation of neuronal communication defines \textit{the payload computational efficiency of an \gls{ANN}, and the computing performance of
	its nodes has only marginal importance}.
The lengthy queueing also leads to irrealistic timings: as discussed in section~\ref{sec:QuantalTime}, 
some physically (not biologically or logically) delayed signals must be dropped to provide
a seemingly acceptable computing performance. Another (wrong) approach to solving the same problem
is introducing conditional computation~\cite{ConditionalComputationBengion:2016}  as discussed in
section~\ref{sec:trainignANNs}. The origin of the issue is that using a shared medium makes the computing system's temporal behavior much more emphasized, and that \textit{the temporal behavior cannot be compensated using methods developed having a timeless behavior in mind}.

The \gls{ANN} type workload introduces an extra handicap: the neurons must communicate the result of an elementary operation after every single operation; in addition through a single shared bus.
	In Fig.~\ref{fig:Neuronal} the inset shows a simple neuromorphic use case: one input neuron and one output neuron communicate through a hidden layer, comprising only two neurons.
	Fig.~\ref{fig:Neuronal}.A mostly shows \textit{the biological implementation: all neurons are directly wired to their partners}, i.e.,
	a system of "parallel buses" (axons) exists. Notice that the operating time also comprises two "non-payload" times ($T_t$): data input and data output, which coincide with the non-payload time of the other communication party. The diagram displays the logical and temporal dependencies of the neuronal functionality.
	The payload operation ("the computing") can only start after its data is delivered (by the, from this point of view,
	non-payload functionality: input-side communication), and the output communication can only begin when the computing finished. Importantly, \textit{communication and calculation mutually block each other}. 
	Two important points that neuromorphic systems must mimic noticed immediately: i/ \textit{the communication time is an integral part of the total execution time}, and ii/ \textit{the ability to communicate is a native functionality} of the system.
	In such a parallel implementation, \textit{the performance of the system}, measured as the resulting total time (processing + transmitting), \textit{scales linearly with increasing either the non-payload communication speed or the payload processing speed}. 

	Fig.~\ref{fig:Neuronal}.B shows a \textit{technical implementation of  a high-speed shared bus} for communication.
	To the right of the grid, the
	activity that loads the bus at the given time is shown.
	A double arrow illustrates the communication bandwidth, the length of which is proportional to the number of 
	packages the bus can deliver in a given time unit.
	We assume that the input neuron can send its information in a single message to the hidden layer; furthermore, the processing by neurons in the hidden layer both starts and ends simultaneously. However, the neurons must compete for accessing the bus, and only one of them can send its message immediately, the other(s)
	must wait until the bus gets released.
	The output neuron can only receive the message when the first neuron completed its sending.    
	Furthermore, the output neuron must first acquire the second message from the bus, and the processing can only begin after having both input arguments. 
	\textit{This constraint results in sequential bus delays both during non-payload processing in the hidden layer and payload processing in the output neuron}.
	Adding one more neuron to the layer introduces one more delay.
	
	At this point, two wrong solutions can be taken: either the second neuron
	must wait until the second input arrives (in biology, a spike also carries a synchronization signal, and triggers its integration), or (in "technical neurons", using
	continuous levels rather than pulses, 
	this synchronization facility is omitted) changes its output continuously,
	as the inputs arrive, and its processing speed enables. 
	In the latter case, however, \textit{until the second input arrives (and gets processed)
		the neuron provides an output signal, differing from the one
		expected based on the mathematical dependence}. As discussed in detail
	in~\textbf{\cite{VeghTemporal:2020}}, this, temporarily may be wrong,
	output signal is known in the electronics, and those "glitches" are eliminated via using a "worst-case" delay for the output signal.
	However, including a serial bus in that computation
	would enormously prolong the needed "worst-case" delay.
	
	Using the formalism introduced above, $T_t=2\cdot T_B + T_d + X$, i.e., 
	the bus must be reached in time $T_B$ (not only the operand delivered to the bus, but also waiting for arbitration: the right to use the shared bus), twice, plus the physical delivery  $T_d$ through the bus.
	The $X$ denotes "foreign contribution": if the bus is not dedicated for "neurons in this layer only", any other traffic also loads the bus: both messages from different layers and the general system messages may make processing slower (and add their contribution to faking the imitated biological effect).
	
	Even if only one single neuron exists in the hidden layer, it must use the mechanisms of sharing the bus, case by case. The physical delivery to the bus takes more time than a transfer to a neighboring neuron (both the arbiter and the bus are in $cm$ distance range, meaning several $nsec$ transfer times, while the direct transfer between
	the connected gates may be in the $psec$ range).
	If we have more neurons (such as a hidden layer) on the bus and work in parallel, they must all wait for the bus. 
	The high-speed bus is very slightly loaded when only a couple of neurons are present. Its load increases linearly with the number of neurons in the hidden layer (or, maybe, all neurons in the system).
	The temporal behavior of the bus, however, is different. 

	Under a biology-mimicking workload, 
the second neuron must wait for all its inputs originating in the hidden layer.
If we have $L$ neurons in the hidden layer,
the transmission time of the neuron behind the hidden layer is $T_t=L\cdot 2\cdot T_B + T_d +X$. 
This temporal behavior explains why "\textit{shallow networks with many neurons per layer \dots scale worse than deep networks with less neurons}"\cite{DeepNeuralNetworkTraining:2016}:
\textit{the physical bus delivery time $T_d$, as well as the processing time $T_p$,
	become marginal if the layer forces to make
	many arbitrations to reach the bus}: the number of the 
neurons in the hidden layer defines the transfer time.
In deeper networks, the system sends its messages at different times in its different layers
(and, even they may have independent buses between the layers), although \textit{the shared bus persists in limiting the communication}.
Notice that there is no way to organize the message traffic: only one bus exists.

At this point comes into picture the role of the workload on the system: the two neurons in the hidden layer want to use the single shared bus, at the same time, for communication. As a consequence, 
\textit{the apparent processing time is several times higher, than the physical processing time,
	and it increases linearly with the number of neurons in the hidden layer} (and, maybe,	with also the total number of neurons in the system, if a single high-speed bus is used).

The ratio of the time spent with forwarding data on the high-speed bus gradually decreases as the system's size increases. 
\textit{In vast systems, especially when attempting to mimic neuromorphic workload, 
	the speed of the bus is getting marginal}.
Notice that the times shown in the figure are not proportional: 
the (temporal) distances between cores are in the several picoseconds range,
while the bus (and the arbiter) are at a distance well above nanoseconds, so \textit{the actual temporal
	behavior (and the idle time stemming from it) is much worse than the figure suggests}. \textit{"The idea of using the popular shared bus to implement the communication medium is no longer acceptable, mainly due to its high contention."}~\cite{ReconfigurableAdaptive2016}.
The extraordinary workload of AI, makes it much harder to operate the systems.

\subsection{The "quantal nature of computing time"\label{sec:QuantalTime}}
One of the famous cases demonstrating existence and competition of those limitations in the fields of \gls{AI} is the research published in~\cite{NeuralNetworkPerformance:2018}.
The systems used in the study were a \gls{HW} simulator~\cite{SpiNNaker:2013} explicitly designed to simulate $10^9$ neurons ($10^6$ cores and $10^3$ neurons per core) and many-thread simulation running on a supercomputer~\cite{SpikingPetascale2014} able to simulate $2\cdot10^8$ neurons (the authors mention $2\cdot10^3$ neurons per core and supercomputers having $10^5$ cores), respectively.
The experience, however, showed~\cite{NeuralNetworkPerformance:2018} that scaling stalled at
$8\cdot 10^4$ neurons, i.e., about four orders of magnitude less than expected.
\textit{They experienced stalling about the same number of neurons,
	for both the \gls{HW} and the \gls{SW} simulator}.

Given that supercomputers have a performance limit~\textbf{\cite{VeghHowMany:2020}},  one can comprehend the former experience:
the brain simulation needs massive communication
(the authors estimated that $\approx 10\%$ of the execution time was spent with non-payload activity), that sharply decreases their
achievable performance, so their system reached the maximum payload performance that their $(1-\alpha)$ enables: the sequential portion was too high.
But why the purpose-built brain simulator cannot reach its maximum expected performance? Is it just an accident that they both stalled at the same value, or some other limiting factor came into play? Paper~\textbf{\cite{VeghBrainAmdahl:2019}} gives the detailed explanation.

The short reply is that digital systems, including brain simulators, have a central clock signal representing an inherent performance limit: no action in the system can happen in a shorter time.  The total time divided by the clock period's length defines maximum performance gain~\textbf{\cite{VeghHowMany:2020}} of a system. If the size of the clock period is the commonly used $1~ns$, and measurement time
(in the case of supercomputers) is in the order of several hours,
clocking does not mean a limitation.

Computational time and biological time are not only not equal, but they are also not proportional. To synchronize the neurons periodically, a "time grid", commonly with $1~ms$ integration time, was introduced.
The systems use this grid time to put the free-running artificial neurons back to the
biological time scale, i.e., they act as a
clock signal: simulation of the next computation step
can only start when this clock signal arrives.
This action is analogous with introducing a clock signal for executing machine instructions: the processor, even when it is idle, cannot begin the execution of its next machine instruction until this clock signal arrives.
That is, \textit{in this case, the clock signal is $10^6$ times longer than the clock signal of the processor}.
Just because neurons must work on the same (biological) time scale, when using this method of synchronization,
the (commonly used)  1~millisecond "grid time" % ("the quantum of computing time")
has a noticeable effect on the payload performance.\footnote{This periodic synchronization shall be a limiting factor in large-scale utilization of processor-based artificial neural chips~\cite{IntelLoihi:2018,TrueNorth:2016}, although thanks to their ca. thousand times higher "single-processor performance", only when approaching the computing capacity of (part of) the brain; or when the simulation turns to be communication-bound.}

The brain simulation measurement~\cite{NeuralNetworkPerformance:2018} enables us to guess the efficacy of
\gls{ANN}s. Given that using more cores only increased the \textit{nominal} performance
(and, correspondingly, its power consumption) of their system, the authors decided
to use only a small fragment of their resources, %for example
only 1\% of the cores available in the \gls{HW} simulator.
In this way,  we can place the efficiency of brain simulation on a supercomputer benchmarking scale, see Fig.~\ref{fig:RooflineBrain}. Under those circumstances,
as witnessed by Fig.~\ref{fig:RooflineBrain}, a performance gain about $10^3$ was guessed for brain simulation. 
Notice that the large-scale supercomputers use about 10\% of their cores in \gls{HPCG} measurement, see also Fig.~\ref{fig:EffDependence2020Log}.
The difference in the efficiency values of \gls{HPL}, \gls{HPCG}, and the Brain comes from the different workloads, which is the reason for the issue.
For \gls{ANN}s the efficiency can be similar to that of brain simulation, somewhat above the
performance gain of \gls{HPCG}, because the measurement time is shorter than that of  \gls{HPCG} on supercomputers, so the corresponding inherent non-parallelizable portion is higher.

Recall also the "communicational collapse" from the previous section: even if communication packages are randomized in time, it represents a colossal peak traffic, mainly if a single global (although high speed) bus is used.
This effect is so strong in large systems that emergency measures must have been introduced, see section~\ref{sec:trainignANNs}.    
In smaller \gls{ANN}s, it was found~\cite{NeuralNetworkPerformance:2018}, that \textit{only a few dozens of thousands of neurons can be simulated on processor-based brain simulators}. This experience includes both many-thread software simulators and a purpose-built brain simulator\footnote{Despite this, Spinnaker2, this time with 10M processors is under construction~\cite{SpiNNaker2:2018}}. Recall also from~\cite{DeepNeuralNetworkTraining:2016}, that "\textit{strong scaling is stalling after only a few dozen nodes}".
For a discussion on the effect of serial bus in \gls{ANN}s, see~\textbf{\cite{VeghTemporal:2020}}.

\subsection{Rooflines of ANNs}
As all technical implementations, computing also has technological limitations.
The "roofline" model~\cite{WilliamsRoofline:2009} successfully describes, that until some needed resource exceeds its technical limitation, utilization of that resource shows a simple linear dependency. Exceeding that resource is impossible: usage of the resource is stalling at the maximum possible level; the two lines form a "roofline". 
In a complex system, such as a computing system, the computing process 
uses different resources, and under other conditions, various resources may 
dominate in defining the "roofline(s) of computing process", see Fig.~\ref{fig:RooflineBrain}.
An example is running benchmarks \gls{HPL} and \gls{HPCG}: as discussed in~\textbf{\cite{VeghHowMany:2020}},
either computing or interconnection dominates the payload performance.

As section~\ref{sec:QuantalTime} discusses, in some cases, a third competitor can also appear on the scene, and even it can play a significant role. That is, it is not easy at all to describe an \gls{ANN} system in terms of the "roofline"~\cite{WilliamsRoofline:2009} model: depending on the actual conditions, the dominant player (the one that defines the top level of the roofline) may change.
Anyhow: it is sure that the contribution of the component, representing the lowest roofline, shall dominate. Still, the competition of parts may result in unexpected issues (for example, see how computation and interconnection changed their dominating rule, in~\textbf{\cite{VeghHowMany:2020}}).
Because of this, Fig.~\ref{fig:RooflineBrain} has limited validity.
It provides, however, a feeling that 1/ for all workflow types a performance plateau exists and already reached; 2/ what value of payload performance gain value can be achieved for different workloads; 3/ where the payload efficiency of the particular kinds of \gls{ANN}s, brain simulation on supercomputers, are located compared to those of the standard benchmarks (a reasoned guess).

\subsection{Role of parameters of computing components}

As discussed in~\textbf{\cite{VeghHowMany:2020}},
\textit{different components of computing systems mutually block each other's operation}.
Datasheet parameters of components represent a hard limit,
valid for ideal, stand-alone measurements, where the utilization is 100\%.
When they must cooperate 
with other components, the way as they cooperate (aka the workload of the system)
defines their soft limit (degrades utilization of units): until its operand(s) delivered, computing cannot start; until computed, result transmission cannot start, mainly when several computing units compete for the shared medium.  	
This competition is the reason why \gls{ANN}s represent a very specific workload,
where weaknesses of principles of computing systems are even more emphasized. 	

\subsection{Training ANNs\label{sec:trainignANNs}}
One of the most shocking features of \gls{ANN}s is their weeks-long
training time, even for (compared to the functionality of brain) simple tasks. The mathematical methods, of course, do not
comprise time-dependence, the technical implementation of 
\gls{ANN}s, however, does: as their time-dependence 
is discussed in details in~\textbf{\cite{VeghTemporal:2020}},
the delivery times of new neuronal outputs (that serve as new 
neuronal inputs at the same time) are only loosely coupled:
the assumption that producing an output means at the same time producing 
an input for some other neuron, works  only in the timeless "classic computing" (and in biology using parallel axons),
see discussing the temporal behavior of the serial bus in~\textbf{\cite{VeghTemporal:2020}}.

To comprehend what change of considering temporal behavior means,
consider the temporal diagram of a 1-bit adder in~\textbf{\cite{VeghTemporal:2020}}.
When using adders, we have a  fixed time when we read out the result.
We are not interested in "glitches", so we set a maximum time until all bits relaxed, and (at the price of losing some performance),
we will receive the final result only; the adder is synchronized.

The case of \gls{ANN}s, however, is different. In the adder, outputs of bit $n$ are the input of bit $n+1$ (there is no feedback), and the bits are wired directly.
In \gls{ANN}s, the signals are delivered via a bus, and the interconnection type and sequence depends on many factors (ranging from the kind of task to the actual inputs). During training \gls{ANN}s,
their feedback complicates the case.
The only fixed thing in timing is that the neuronal input arrives inevitably only after a partner produced it. The time ordering of  delivered events, however, is not sure:
it depends on technical parameters of delivery, rather than on the logic that generates them.
Time stamping cannot help much. There are two bad choices. Option one is that neurons should have a (biological) sending time-ordered input queue and begin processing when all partner neurons have sent their message. That needs  a synchronous signal and leads to 
severe performance loss (in parallel with the one-bit adder).
Option two is that they have a (physical) arrival-time ordered queue, and they are 
processing the messages as soon as they arrive. This technical solution enables us to give feedback to a neuron 
that fired later (according to its timestamp), and set a new neuronal variable state; which
is a "future state" when processing a message received physically later, but with
a timestamp referring to a biologically earlier time.
A third, maybe better, option would be to maintain a biological-time ordered queue,
and either in some time slots (much shorter than the commonly used "grid time") send out
output and feedback, individually process the received events, and send back feedback and output immediately.
In both cases, it is worth to consider if their effect is significant (exceeds some tolerance level compared to the last state)
and mitigate the need for communication also in this way.

We start showing an input during training, and the system begins to work, using the synaptic weights valid before showing that input. Those weights may be randomized, or maybe that they correspond to the previous input data. The signals that
the system  sends, are correct, but a receiver does not know the future: it processes
a signal only after it was physically delivered\footnote{Even if the message envelope contains a time stamp}, meaning that it (and its dependents) may start to adjust their weights to 
a state that is still undefined. In~\textbf{\cite{VeghTemporal:2020}}, the first AND gate has quite a short
indefinite time, but the OR has a long one.

When playing chess against its opponent, a faster computer can be advantageously used to analyze the forthcoming moves.
Even it can compute all future moves before its opponent makes the next move. However, with publishing its next move, it must wait the next move made by its opponent, otherwise it may publish a wrong move. Sending the feedback to its opponent as soon as the next move is computed --i.e., without synchronization-- results in a quickly computed but maybe wrong move. Without synchronization, the faster is the computer, the worse is its performance as chess player. 

At the beginning of their operation, some component neurons of the network may have undefined states and weights.
Their operation is essentially an iteration, where --without synchronization-- \textit{the actors mostly use mostly wrong input signals,
	and surely adjust their weights to false signals initially and with significant time delay at later times}. If we are lucky (and consider that we are working with unstable states in the case of more complex systems), the system will converge,
but painfully slowly. Or not at all. \textit{Not considering the temporal behavior of the network leads to painfully slow and doubtful convergence.}

\textit{Synchronization is a must, even in \gls{ANN}s. We must take care when using accelerators, feedback and recurrent networks. The time matters.}
Computing neuronal results faster, to provide feedback more quickly, 
cannot help much, if at all. Delivering feedback information also needs time and uses the same shared medium,
with all its disadvantages. In biology, the "computing time" and the "communication time" are in the same order of magnitude. In computing, the communication time is very much longer than computation, that is,
\textit{the received feedback refers to a time} (and the related state variables) \textit{that was valid a very long time ago}.
In biology, spiking is also a "look at me" signal: the feedback shall be sent to \textit{that} neuron,
reflecting the change its output caused\footnote{See the Hebbian learning: the neuron uses its inputs and output, exclusively.}. Without this, neurons receive feedback about "the effect of all fellow neurons, including me". Receiving a spike defines the time of the beginning of signal's validity; "leaking" also defines their "expiration time". When using spiking networks, their temporal behavior is vital.

In excessive systems,
some result/feedback events must be dropped because of long queuing to provide seemingly higher performance. The logical dependence that the feedback is computed from the results of the neuron that receives the feedback,
the physical implementation of the computing system converts to time dependence~\textbf{\cite{VeghTemporal:2020}}.
Because of this time sequence, the feedback messages will arrive at the neuron later (even if at the same biological time, according to their time stamp they carry),
so they stand at the end of the queue. Because of this, it is highly probable that they "\textit{are dropped if the
	receiving process is busy over several delivery cycles}"~\cite{NeuralNetworkPerformance:2018}. 
In vast systems, the feedback in the learning process involves results based on undefined inputs,
and the calculated and (maybe correct) feedback may be neglected.

An excellent "experimental proof" of the claims above is provided in~\cite{ConditionalComputationBengion:2016}. With the words of that paper:
"\textit{Yet the task of training such networks remains a challenging optimization problem. Several related problems arise: very long training time (several weeks on modern computers, for some problems), the potential for over-fitting (whereby the learned function is too specific to the training data and generalizes poorly to unseen data), and more technically, the vanishing gradient problem}". 
"\textit{The immediate effect of activating fewer units is that propagating information through the network will be faster, both at training as well as at test time.}"
This effect also means that the computed feedback, based maybe on undefined inputs, reaches the previous layer's neurons faster.
A natural consequence is that (see their Fig.~5): "\textit{As $\lambda_s$ increases, the running time decreases, but so does performance.}"	
Similarly, introducing the spatio-temporal behavior of ANNs, even in its simple form, using separated (i.e., not connected in the way proposed in~\textbf{\cite{VeghTemporal:2020}})  time and space contributions to describe them, significantly improved the efficacy of 
video analysis~\cite{SpatiotemporalLearning:2018}.

The role of time (mismatching) is confirmed directly,
via making investigations in the time domain. "\textit{The CNN models are more sensitive to low-frequency channels than high-frequency channels}"~\cite{LearningFrequencyDomain:2020}:
the feedback can follow the slow changes with less difficulty compared to the faster changes.

	\section{Summary}
The strongly simplified computing paradigm, proposed by von Neumann,  surely has severe limitations when applied to today's technology.
According to von Neumann, it is doubly $unsound$ if one attempts to mimic neural operation based on a paradigm that is $unsound$ for that goal, on a technological base (other than vacuum tubes) for which the paradigm is $vitiated$. As predicted by Amdahl, 
large (many-processor) machines have inherent disadvantage in computing. Hennessy pointed out that the efficiency of
distributed systems is strongly limited and heavily depends
on their workload. 

The operating characteristics of \gls{ANN}s are practically unknown, mainly because of their mostly proprietary design/documentation.
We reviewed some general features
of \gls{ANN}s, with the goal to provide help in designing
new systems, and to understand their scaling behavior. The existing theoretical predictions 
and measured results 
show good agreement,
but dedicated measurements using well-documented benchmarks and a variety of well-documented architectures are needed. The low efficacy of our designs forces us to change our design methods. On one side, it requires a careful design method when using existing components (i.e., to select the "least wrong" configuration; millions of devices shall work with low energy and computational efficacy!). On the other side, it urges working out a different computing paradigm %~\textbf{\cite{VeghModernParadigm:2019}} 
(and architecture %~\textbf{\cite{VeghSPAEMPA:2020}}
based on it).

%	In his famous "First Draft"~\cite{EDVACreport1945}, von Neumann formulated: "\textit{6.3 At this point the following observation is necessary. In the human nervous system the conduction
%		times along the lines (axons) can be longer than the synaptic delays, hence our above procedure of
%		neglecting them aside of $\tau$ [the processing time] would be \textbf{unsound}. In the intended vacuum tube interpretation,
%		however, this procedure is justified: $\tau$ is to be about a microsecond, an electromagnetic impulse
%		travels in this time 300 meters, and as the lines are likely to be short compared to this, the conduction
%		times may indeed be neglected. (\textbf{It would take an ultra high frequency device -- $\approx 10^{-8}$ seconds
%		or less -- to vitiate this argument.}})"
%	That is, (at least) \textit{since the processor frequency exceeded 0.1~GHz, it is surely \textit{unsound} to use von Neumann's computing paradigm in its unchanged form, neglecting transmission time.}

%\section*{Declarations}
%
%All manuscripts must contain the following sections under the heading 'Declarations'.
%
%If any of the sections are not relevant to your manuscript, please include the heading and write 'Not applicable' for that section.
%
%To be used for all articles, including articles with biological applications
\subsection*{Funding}
Project no. 136496 has been implemented with the support provided from the National Research, Development and Innovation Fund of Hungary, financed under the K funding scheme

	 \section*{Conflict of interest}
	
	The authors declare that they have no conflict of interest.

\subsection*{Availability of data and material} 
Not applicable

\subsection*{Code availability} (software application or custom code)
Not applicable

\subsection*{Authors' contributions}
Not applicable



%	\bibliographystyle{spmpsci}       % APS-like style for physics
%	\bibliography{../../../CommonBibliography%
%		,../../../CommonPrivateBibliography%
%		,../../../CommonNeuronalBibliography%
%	}

%% Non-BibTeX users please use
%\begin{thebibliography}{}
%%
%% and use \bibitem to create references. Consult the Instructions
%% for authors for reference list style.
%\end{thebibliography}

\end{document}